\documentstyle[prd,aps,floats,epsfig]{revtex}
\begin{document} 
\author{E.J.Copeland$^a$, Jo\~ao Magueijo$^b$ 
 and D.A.Steer$^c$ } 
\address{{\it a})Centre for Theoretical Physics, University of 
Sussex, 
 Brighton BN1 9QH, U.K.\\ 
{\it b}) Blackett Lab., Imperial College, Prince Consort Road, London 
SW7 2BZ, U.K.\\ 
{\it c}) D.A.M.T.P., Silver Street, Cambridge, CB3 9EW, U.K.} 

\title{Cosmological parameter dependence in local string theories of 
structure formation}

\maketitle  
\vspace{1cm}
\begin{abstract} 
We perform 
a new and accurate study of the 
dependence on cosmological parameters of structure formation with 
local cosmic strings. The crucial new ingredients are the inclusion 
of the effects of gravitational backreaction on the evolution of the 
network, and the  accurate evolution of the network through 
the radiation to matter transition. Our work re-iterates the fact 
that expanding Universe numerical simulations only 
probe a transient regime, 
and we incorporate our results into the unequal time correlators  
recently measured. We then compute the CMB and CDM  
fluctuations' power spectra for various values of the  Hubble constant  
$H_0$ and baryon fraction $\Omega_b$. We find that, whereas 
the dependence on $\Omega_b$ is negligible, 
there is still a strong dependence on $H_0$. 
\end{abstract} 
\date{\today} 


\newcommand{\gsim}{\raise.3ex\hbox{$>$\kern-.75em\lower1ex\hbox{$\sim$}}} 
\newcommand{\lsim}{\raise.3ex\hbox{$<$\kern-.75em\lower1ex\hbox{$\sim$}}}

\newcommand{\hatC}{\hat{C}}

\newcommand{\paa}{\partial} 
 
\newcommand{\bxi}{\bar{\xi}} 
\newcommand{\bgamma}{\bar{\gamma}} 
 
\newcommand{\GGmu}{\Gamma G \mu} 
\newcommand{\half}{{\frac{1}{2}}}


\renewcommand{\thefootnote}{\arabic{footnote}} 
\setcounter{footnote}{0} 
\typeout{--- Main Text Start ---} 


\section{Introduction}
Modern theories of structure formation are based on two 
paradigms: cosmological inflation \cite{infl} and  
topological defects \cite{csreviews}. Inflationary scenarios produce  
structure from vacuum quantum fluctuations, and 
have been well studied.  Hence upcoming high precision 
experiments aimed at measuring 
the Cosmic Microwave Background (CMB) anisotropy and Large
Scale Structure (LSS) power spectra should also provide  
information on the model's free parameters.  Parameters internal to the
theory are, for example, the primordial tilt, the 
perturbations' amplitude, or the percentage  
of tensor modes. 
Others describe the current state of the Universe, and include 
the ratios to the critical density, of the matter density, $\Omega$, 
of the vacuum energy,  $\Omega_\Lambda$, of the baryon density, $\Omega_b$,  
and the Hubble constant $H_0$.

Progress in defect theories has been slower.  Here, as the  
Universe cools down, high temperature 
symmetries are spontaneously broken, and remnants of the unbroken phase 
may survive the transition \cite{csreviews}.  These ``vacuum'' defects later  
seed fluctuations in the CMB and LSS. The defect evolution is  
highly non linear, thereby complicating the computation of these 
fluctuations. However, recent breakthroughs in method and 
computer technology have allowed unprecedented progress  
\cite{Neil,James1,CHM,avelino},
although the results for a standard combination of parameters ($\Omega=1$, 
$\Omega_\Lambda=0$, $\Omega_b=0.05$, $H_0=50$ km/sec/Mpc) 
were discouraging. Tuning parameters to fit the data has not been  
thoroughly attempted, although initial indications
are that it is possible to improve the fits considerably
\cite{avelino,James2}.  Also, combinations of 
strings and inflation with the standard parameters seem to 
fit the data much better than each separate component  
\cite{CHM2,stinf1,stinf2}.

Unlike inflation, defect theories only have one internal  
parameter: the symmetry breaking energy scale (which is also the 
string mass per unit length $\mu$ for local cosmic 
strings). Nonetheless, quantities like $\Omega$, $\Omega_b$, 
$\Omega_\Lambda$, or $H_0$ are also free parameters in defect theories, 
although it has been suggested in the context of global theories 
that the dependence on these is much weaker 
\cite{Neil}. The purpose of this paper is to reexamine this statement, 
in the context of local cosmic strings.  
 
Before such a study can be attempted a number of aspects of the work on 
strings \cite{James1,CHM,avelino} need to be refined.  
The effect of the radiation to matter (RM) transition, and the transition to 
curvature or vacuum domination upon the network need to be established,  
if sensitivity to the cosmological parameters is to be fully considered.  
However, these effects were neglected in 
the recent highest accuracy calculations of structure formation with  
local cosmic strings \cite{CHM}, which make use of unequal time 
correlators (UETC). Here we will use the  
3-scale model of Austin, Copeland and Kibble \cite{ACK,ACK2,EdRome}  
to evaluate those effects in a flat Universe with no cosmological  
constant.  A further key ingredient of this model is the inclusion of 
gravitational backreaction (GBR). These results 
are incorporated into the UETCs of \cite{CHM}, and 
the CMB and LSS power spectra are obtained on a grid of values for $\Omega_b$ 
and $H_0$. We argue that this combination of techniques provides 
the most accurate calculation of these spectra  
allowed by current computer technology.

This paper is set up as follows.  First, in section \ref{threescale},
we discuss the aspects of the 3-scale model which are relevant 
to this work.  In section \ref{method} we then explain how these results
are incorporated into the UETCs of \cite{CHM}.  The resulting CMB and
LSS power spectra, and their dependence on $\Omega_b$ 
and $H_0$ are then presented in section \ref{results}.  Conclusions
are given in section \ref{conc}.

\section{The 3-scale model}
\label{threescale}

The 3-scale model \cite{ACK,ACK2} provides an analytic 
description of the string network in terms 
of three physical lengths $\xi$, $\bxi$ and $\zeta$.    
The mean string separation is $\xi$ so that the energy  
density in strings is $\rho = \mu / 
\xi^2$.  (Renormalisation effects are encoded in 
$\xi$ and not $\mu$.)  The mean correlation length  
along the strings is  
$\bxi (\neq \xi)$.  Finally the 
small scale structure is described by $\zeta$ which is  
effectively the interkink distance.  
These scales are implicitly functions of each other, of time $t$ and of 
parameters describing the effects of expansion, gravitational 
radiation, GBR, loop formation and intercommutation.  They also
determine the root mean square string velocity $v_{rms}$.
In this paper we have solved the evolution equations 
for $d\xi/da$, $d\bxi/da$ and $d\zeta/da$ 
in terms of the scale factor $a$ and the variables 
$\gamma = 1/H \xi$, 
$\bar{\gamma} = 1/H \bxi$ and $\epsilon = 1/H \zeta$ where $H$ is the 
Hubble parameter.  We do not write out these rather long equations
here:  they follow directly from those given in \cite{ACK} by a simple
change of variables from $t \rightarrow a$.
Of particular interest is the effect of GBR, 
the behaviour of the network across the 
RM transition, and the question of whether or not the system has reached 
a scaling regime ($\epsilon$, $\gamma$ and $\bgamma$ all
constant) today. 

The 3-scale model could be criticised for 
containing many undetermined parameters.  However, 
some of these ---such as the  
infinite string intercommutation probability $\chi$, 
the chopping efficiency $c$ which gives the rate of loop 
formation, and 
$\GGmu$ the rate of gravitational radiation, also appear in 1-scale 
models, for example \cite{CJAP}.  Other parameters are 
tightly constrained by 
numerical simulations of cosmic string evolution (see below).  Indeed, 
the only undetermined parameters are $k$, which 
describes the excess small-scale kinkiness on loops compared to long 
strings, 
and $\hat{C}$ which  
determines the rate at which GBR smooths 
the small-scale kinkiness; $\left. d\zeta / dt \right|_{GBR} =  
\hat{C} \GGmu  $.  Neither quantity is to be found in current 
numerical simulations nor in other analytic models.  
One further point; we assume here 
that gravitational radiation and not particle production is the  
dominant source of energy loss \cite{Mark}.  This implies that
$\hat{C} > \hat{C}_{crit} > k$ for $\hat{C}_{crit}$ a given constant 
of order 1 \cite{ACK}.  The reader is referred to \cite{PP} for a
recent discussion of the effect of different decay mechanisms on 
the observational consequences of strings.
\begin{figure}  
\centerline{\epsfig{file=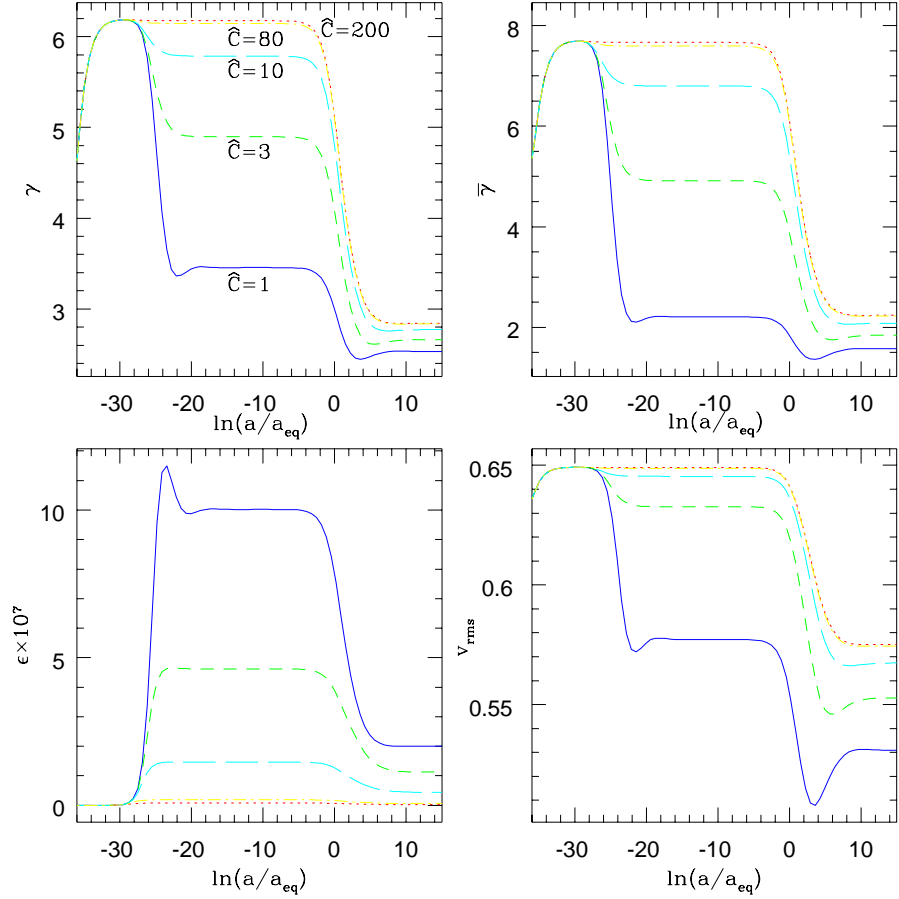,width=13 cm,angle=0}}
\caption{Evolution of $\gamma$, $\bar{\gamma}$, $\epsilon$ and
$v_{rms}$ for different values of
$\hat{C}$ and $H_0 = 50$ km/sec/Mpc.  Other parameters are
 $k=0.05$, $\chi=0.1$, $c=0.15$ with $\GGmu = 10^{-8}$ to emphasise the 
transient scaling regime (see text).} 
\label{string1} 
\end{figure} 
\begin{figure}  
\centerline{\epsfig{file=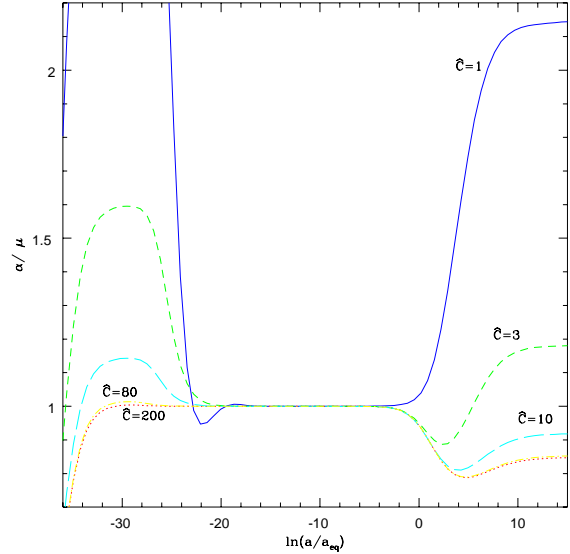,width=8 cm,angle=0}}
\caption{Plot of $\alpha$, the 
comoving energy density in strings multiplied by the
conformal time squared ($\alpha = N \rho_{c}\tau^2 = N \mu H^2 \tau^2 a^2
\gamma^2$ where the normalisation $N$ is such that $\alpha=1$ deep in
the radiation era).  Different curves correspond to different 
values of $\hat{C}$ and other parameters are as in the previous caption.} 
\label{string2} 
\end{figure} 

Some important features of the solutions for $\gamma$, $\bgamma$,
$\epsilon$ and $v_{rms}$ are shown in 
Fig.~\ref{string1} for different values of $\hat{C}$.  
One of these is that there are two 
different scaling solutions.  The first, `transient' 
solution, has constant values of $\gamma_{tr}$, 
$\bgamma_{tr}$ and $v_{rms}^{tr}$ ({\em independent} 
of $\hat{C}$), but $\epsilon$ is growing: small scale structure  
is building up on the network. 
We believe that this corresponds to 
the scaling solutions of numerical simulations which have no GBR (so 
$\hat{C}=0$).     
From the evolution equations one can obtain expressions for  
$\gamma_{tr}$, $\bgamma_{tr}$ and $v_{rms}^{tr}$ \cite{ACK} and thus 
bound them by data from simulations, 
which in turn 
constrains many of the 3-scale model parameters  
as claimed above.  
We comment that the {\em duration} 
of the transient regime {\em does} depend on  
$\hat{C}$, ending when $\epsilon \sim 1/\GGmu \hat{C}$.  
Thus if, as in numerical simulations, we were 
to set 
$\hat{C}=0$ it would last indefinately, with 
$\epsilon \rightarrow \infty$ eventually  
causing the strings to disappear ($\gamma, \bgamma \rightarrow 0$).
Finally, recall that in Ref~\cite{Mark} particle
production was found to be the dominant source of energy loss from the 
network. 
The analogue here correpsonds 
to having $k > \hat{C}_{crit}$ for which 
the evolution equations lead to a scaling regime  
with $\gamma \sim \epsilon \sim \bgamma$; a 1-scale model \cite{EdRome}.   

Fig.~\ref{string1} shows that following the transient 
regime is a new `full' scaling regime in which {\em all} the 
variables reach new constant values, apart from a period between  
$-5 \lsim \ln(a/a_{eq}) \lsim 10$ (including recombination and even
today) resulting from the RM transition.  It appears that the string
network might only just, if at all, be reaching a scaling solution today.
Indeed, across the transition, different scales seem to reach scaling 
at different rates which are also dependent on $\hat{C}$.    
We also see that as $\hat{C}$ 
increases $\zeta$ increases since kinks  
are smoothed out more effectively; $v_{rms}$ is also increased. 
It is this full scaling solution, and not the transient one that 
is clearly the relevant one for structure 
formation.  However, we noted above that $\hat{C}$ is so far unknown.  
We have therefore chosen to work in the limit 
where the scaling solutions are as insensitive as 
possible to $\hat{C}$.  From Fig.~\ref{string1}  
this corresponds to large values of 
$\hat{C}$ for which the decrease in physical energy density across the 
radiaton-matter transition is also largest. 
In that limit, the solutions plotted in 
Fig.~\ref{string1} are also insensitive to the precise value 
of $k (< \hat{C}_{crit})$---the 
exception is $\epsilon$ which decreases as `particle production' ($k$) 
increases.

\section{Method of determination of the Powerspectra}
\label{method}

We have used the full scaling solutions including GBR with large 
$\hat{C}$ to update the results of \cite{CHM}, where 
the power spectra in string induced CMB fluctuations ($C_\ell$) 
and LSS ($P(k)$) were computed following methods first proposed  
in \cite{Neil}. 
Of key importance is to obtain 
from simulations the 2-point functions of the defect stress-energy  
tensor $\Theta_{\mu\nu}$.  These so-called unequal time correlators
(UETC) are defined as 
$ 
{\langle \Theta_{\mu\nu}({\bf k},\tau)\Theta^\star_{\alpha\beta}  
({\bf k},\tau')\rangle}\equiv{\cal C}_{\mu\nu,\alpha\beta}(k,\tau,\tau ')  
$ 
where ${\bf k}$ is the wavevector and $\tau$ and $\tau '$   
are any two (conformal) times.
UETCs contain all the information required for computing the 
power spectra in  defect induced fluctuations.   
They fall off sharply away from the equal-time diagonal ($\tau=\tau'$), 
and are tightly constrained by requirements of self-similarity   
(or scaling) and causality. These properties allow extrapolation of 
measurements made over the limited dynamical range of simulations,  
thus giving the defect history over the whole life of the Universe.

However, it is clear that scaling breaks down near the RM
transition (Fig.~\ref{string1}). This may be incorporated into the UETCs
by measuring radiation and matter epoch UETCs, transition UETCs, 
and then interpolating \cite{Neil}.  Such a procedure was applied to 
global defects, but cannot be followed with the simulations of \cite{CHM}, 
in which the expansion damping effects upon the network are neglected. 
Nonetheless these simulations are still the only available complete source 
of local string UETCs (see however \cite{avelino,wu}). The 3 scale 
model of \cite{ACK}, while allowing the most rigorous study to date 
of departures from scaling, does not provide sufficient information 
for constructing UETCs.  
 
\begin{figure}  
\centerline{\epsfig{file=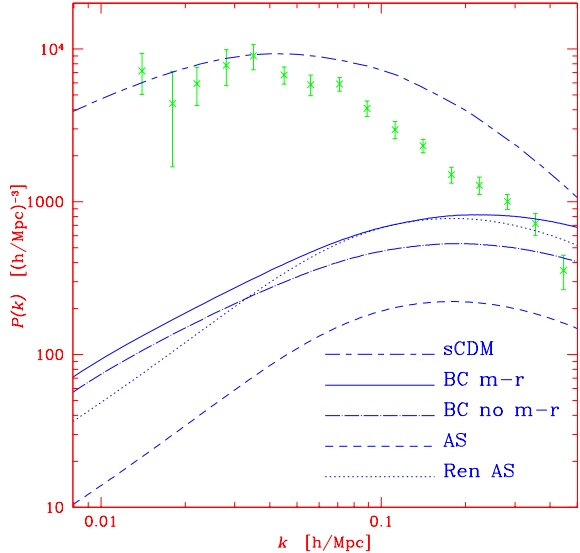,width=8 cm,angle=0}}  
\caption{The CDM power spectrum in sCDM scenarios (top curve), 
and local cosmic strings with large $\hat{C}$. 
We consider the result of a Boltzmann 
code calculation, with and without the effects upon the network 
of the RM transition included, and also the 
results of a Albrecht and Stebbins approximation. The latter 
underestimates the normalization and 
distorts the shape of $P(k)$.} 
\label{fig3}  
\end{figure} 
We therefore decided to adopt a hybrid approach, and incorporate
the information supplied by the 3-scale model of \cite{ACK} into 
the UETCs of \cite{CHM}. The UETCs are renormalised 
with a non-scaling factor describing the change in  
comoving string density 
across the RM transition, as determined from the  
3-scale model.  That is we make the replacement
\begin{equation}
{\cal C}_{\mu\nu,\alpha\beta}(k,\tau,\tau ')  
\rightarrow{\cal C}_{\mu\nu,\alpha\beta}(k,\tau,\tau ')  
\alpha(\tau)\alpha(\tau ')
\label{rep}
\end{equation}
where $\alpha(\tau)$ measures the change in the defect scaled comoving
density.  More specifically, $\alpha(\tau) = N \tau^2 \rho_{c}=
N \mu (H a \tau \gamma)^2$ where $N$ is a normalisation factor 
chosen such that $\alpha(\tau)=1$ deep in the radiation era, and
$\rho_c$ is the comoving energy density in srings.  Note
that $\alpha$ depends on $\gamma$ and hence (see figure \ref{string1}) 
on $\hat{C}$:  this $\hat{C}$ dependence of $\alpha$ is plotted in figure
\ref{string2}.  The procedure given in equation (\ref{rep}) 
was also followed in \cite{James2},  
in the context of a semi-analytical model for string induced  
fluctuations. The rationale lies on the fact 
that UETCs are proportional to the square of the defect comoving density, 
if one assumes no correlation between individual defects (as in 
the model of \cite{James1}). The presence of inter-defect correlations,  
non-negligible for small wavelengths, means that the shape of the  
correlators could in principle also change with $\gamma$.  However, it
is difficult to extract this information from the 3-scale model, and
we expect the effect upon the CMB and LSS to be small. Naturally
all UETCs (scalar, vector, and tensor) are multiplied by the same 
time-dependent factor. This then propagates to the various eigenmodes. 

It is crucial to notice that it is the comoving density, and not the  
physical density that enters the above scaling argument. The 
conversion factor between physical and comoving densities changes  
appreciably from the radiation to the matter epoch.  Indeed for large
$\hat{C}$ it accounts for most of the drop in defect physical density, 
and as a result the strings' comoving density (or $\alpha(\tau)$ plotted 
in figure \ref{string2}) drops only by about 20\% for large $\hat{C}$.
For small $\hat{C}$ the comoving energy density actually {\em increases} as
can be seen from figure \ref{string2}.
 
We reran the codes used in \cite{CHM} with these modifications
(working throughout with large $\hat{C}$), and with two purposes 
in mind: to compare the effects of neglecting the RM transition with   
approximations used by other groups, and to survey parameter space in cosmic 
string models.

\section{Results}
\label{results}

We found that the effect of neglecting the RM 
transition is not very large, and certainly much 
smaller than the errors induced by using the Albrecht and Stebbins (AS) 
approximation \cite{ALST} (or a variation thereof 
\cite{avelino,cheung}).  
As suggested in \cite{CHM}, $G\mu$ normalized to COBE  
increases by a factor of the same order as the decrease in string density 
in the matter epoch. With standard parameters  
($\Omega=1$, $\Omega_\Lambda=0$,    
$\Omega_b = 0.05$ and $H_0 = 50$ km/s/Mpc) we 
found that $G\mu$ increases from $1\times 10^{-6}$ to  
$1.2\times 10^{-6}$ \cite{note}.
The COBE normalized $C_\ell$ spectrum, 
in this cosmology, hardly changes at all. The bias in spheres with 
radius 100 Mpc $h^{-1}$ changes from 4.9 to 4.2: the inclusion 
of the RM transition improves the bias problem 
of cosmic strings but not by much. The effect on the shape and normalization 
of $P(k)$ is plotted in Fig.~\ref{fig3}. 
By contrast, using the AS approximation instead of a Boltzmann code  
(or one of its truncations \cite{CHM,cmbfast}) induces 
large uncertainties in normalization and shape of $P(k)$.  (We
obtained the AS curves of Fig.~\ref{fig3} by replacing the
variables $\xi$ and $\chi$ of \cite{ALST} with the $\xi$ and
$\bar{\xi}$ of the 3-scale model respectively.)

\begin{figure}  
\centerline{\epsfig{file=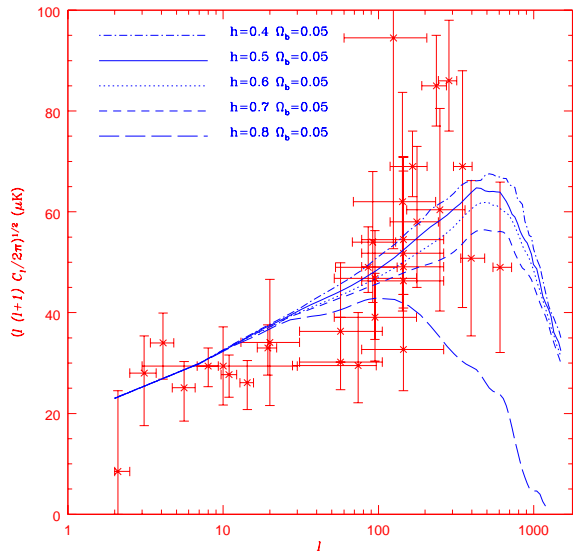,width=8 cm,angle=0}}  
\caption{The CMB power spectra predicted by cosmic strings  
for different values of the Hubble constant.
We have plotted $(\ell(\ell+1)C_\ell/2\pi)^{1/2}$ in $\mu K$,  
and superposed several experimental points. The effect 
of varying $\Omega_b$ is negligible, and so for clarity in this picture 
it was kept fixed. }  
\label{fig4}  
\end{figure}
\begin{figure}  
\centerline{\epsfig{file=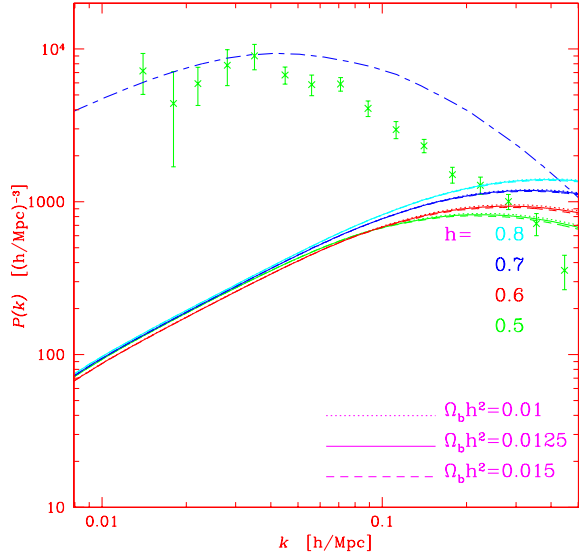,width=8 cm,angle=0}}  
\caption{The effect of changing $h$ and $\Omega_b h^2$ upon 
the power spectrum $P(k)$ in CDM.}  
\label{fig5}  
\end{figure}

  
Although a small effect, only once the RM transition has been properly  
taken into account can we start exploring dependence 
on cosmological parameters. 
We considered $\Omega_b$ and $h$, and  
solved for the $C_\ell$ and $P(k)$ on a grid with $0.4<h<0.8$  
and $0.01<\Omega_b h^2<0.015$. The latter satisfies nucleosynthesis 
constraints, the former accommodates currently accepted ranges of the Hubble 
parameter. In Figs.~\ref{fig4} and \ref{fig5} 
we plot $C_\ell$ and $P(k)$ extracted from 
this grid to illustrate the observed variations. 
The $C_\ell$ spectrum of anisotropies 
induced by strings 
seems insensitive to $\Omega_b$, but depends sensitively 
on $h$ (Fig.~\ref{fig4}). In particular, local cosmic strings, unlike global  
defects, exhibit a Doppler peak, but only if the Hubble constant is not  
too large. The results of \cite{CHM} are therefore not generic, but depend 
on the choice of $h$; indeed they are consistent with the results of
\cite{steb} in which strings 
are directly coupled to a Boltzmann code.  In that work it was found 
that strings do not have a Doppler peak, with $h=0.8$. 
Notice how the dependence of the $C_\ell$ on $h$ is non-linear,
with a much larger jump from $h=0.7$ to $h=0.8$ than, say,
from $h=0.6$ to $h=0.7$.  
The effect of both $h$ and $\Omega_b$ upon the CDM power 
spectrum is very small if $P(k)$ is plotted using  units of $h$/Mpc
for wavenumbers (Fig.~\ref{fig5}). This is in agreement with
small $k$ analytical predictions (eg. \cite{av}), but breaks down
at large $k$ as can be seen by the diverging curves in Fig.~\ref{fig5}.
The string bias problem at 
100 Mpc $h^{-1}$ is clearly not solved by changing $h$ or 
$\Omega_b$ (but see \cite{CHM,avelino,James2,CHM2} for other solutions). 
Finally, recall that we have worked with large $\hat{C}$.  For
small $\hat{C}$'s the strings' comoving energy density increases
across the RM transition and the bias problem is in fact worse.
 
\section{Conclusion}
\label{conc}

To summarize, in this paper 
we have studied the impact of GBR on 
cosmic string network evolution, and properly accounted for 
its properties during the RM
transition.  Not only 
did we suggest that this departure from scaling lasts longer than 
that obtained in other models, but we re-iterated the existence of two 
different scaling solutions.  The full scaling solution is the 
relevant one for structure formation and cannot be observed in 
numerical simulations.   
 
These results, with large $\hat{C}$, were then used to 
update the calculations of  
\cite{CHM}, and cut a first, safe slice through parameter 
space in string scenarios. Dependence on $h$ and $\Omega_b$ was considered, 
the conclusions being that the CMB results are sensitive to $h$ but less 
so to $\Omega_b$. If forthcoming experiments were to 
reveal a $C_\ell$ spectrum without 
secondary Doppler peaks---a tell-tale signal of a defect theory
\cite{inc}---then we would lose ability to determine $\Omega_b$, since 
information on $\Omega_b$ is hidden in the secondary Doppler peaks which 
are generically erased by incoherence, in defect scenarios. 
However, for $h$, our conclusions are more positive. Local strings 
exhibit a Doppler peak whose height we have
demonstrated decreases quite sharply with an increasing Hubble constant.

Larger regions of parameter space require surveying. In particular 
as pointed out in refs \cite{avelino,James2}, 
strings with a cosmological constant 
are a promising combination. However we feel that all treatments,
including our own, become too qualitative to be reliable
if curvature or $\Lambda$ are present.  Considerable analytical and
numerical work
with $\Omega_\Lambda\neq 0$, and in open models, 
is required before a more quantitative discussion
is possible.  The first step in the context of the method we are proposing
here would be to include $\Lambda$ into the 3-scale
model.

\section*{Acknowledgements}  
We thank Andy Albrecht, Anne Davis, Pedro Ferreira,
Mark Hindmarsh, Tom Kibble and 
Neil Turok for useful discussions and comments.  We acknowledge 
financial support from PPARC (E.J.C and D.A.S) and the Royal Society 
(J.M).

\typeout{--- No new page for bibliography ---}


\begin{thebibliography}{99}
\bibitem{note}
We have assumed that the strings decay into gravitational
radiation and relativistic loops.
\end{thebibliography}
\end{document}